\documentclass[11pt,a4paper]{article}
\usepackage{amsfonts}
\usepackage{amssymb}
\usepackage{graphicx}
\usepackage{amsmath}
\usepackage[dvipdfm]{hyperref}
\usepackage{enumerate}
\RequirePackage{mathrsfs} \RequirePackage[sc]{mathpazo}
\RequirePackage{wasysym} \RequirePackage{setspace}

\makeatletter \@addtoreset{equation}{section}
\newcommand{\be}{\begin{equation}}
\newcommand{\ee}{\end{equation}}
\newcommand{\bea}{\begin{eqnarray}}
\newcommand{\eea}{\end{eqnarray}}
\begin{document}

\title{Quiver Gauge theories from   Lie Superalgebras}
\author{  Adil  Belhaj$^{1}$\thanks{belhaj@unizar.es},  M. Brahim Sedra$^{2}$\thanks{msedra@ictp.it}    \\
\\
{\small  $^{1}$Centre of Physics and Mathematics, CPM-CNESTEN, Rabat, Morocco   }  \\
{\small $^{2}$  Universit\'{e} Ibn Tofail, Facult\'{e} des Sciences,
D\'{e}partement de Physique, LHESIR, K\'{e}nitra, Morocco}}
\maketitle

\begin{abstract}
We discuss quiver gauge models with matter fields based  on Dynkin
diagrams  of  Lie superalgebra structures.  We  focus on $A(1,0)$
case and we find  first that it can be related to intersecting
complex cycles with genus $g$.  Using toric geometry, $A(1,0)$
quivers are analyzed in some details and it is  shown  that $A(1,0)$
can be used to incorporate fundamental fields to a product of  two
unitary factor groups. We
 expect that this approach can be applied to other kinds of Lie
superalgebras.\newline KeyWords:   Quiver gauge theories, Lie
superalgebras, Toric geometry.
\end{abstract}

\thispagestyle{empty} \newpage \setcounter{page}{1} \newpage

Four dimensional gauge theories has attracted a special attention in
connection with supergravity  models living in higher dimensional
theories. In particular, they appears in the study of D-brane world
volume gauge theories embedded in such theories  compactified either
on Calabi-Yau manifolds or $G_2$ manifolds. The group and matter
content of the resulting models are obtained from the ADE
singularities of the K3 fibers and the non-trivial geometry
describing the base space of the internal manifolds  resepectively
\cite{KKV,KMV,BFS,ABS1,ABS2,W,BRSS}. In this way, the complete set
of physical parameters of the gauge theory is related to the moduli
space of the associated manifolds. This program is called geometric
engineering allowing to have  exact solutions for  the Coulomb
branch moduli space\cite{KMV}. A nice way to encode the physical
information of such gauge theories is to use the quiver approach
\cite{ACCERV1,ACCERV2}. The corresponding gauge models are usually
called quiver gauge theories. For  a gauge theory with gauge
symmetry given by the product
\begin{equation}
 \label{G}
 G =\prod_{i}G_i,
\end{equation}
 we represent its   physical content by a quiver, or
equivalently a graph, formed by nodes and links.  For each node, one
associates a gauge  factor $G_i$ ( $i$ denotes  the number of
nodes).  Links between two nodes are associates with $a_{ij}$ matter
transforming in bi-fundamental representation of two factor of the
gauge group $G_i \times G_j$.

Recently, many examples of quivers have been built in connection
with string theory moving on manifolds with special holonomy groups.
In particular, such quivers   have been  based on toric graphs and
Dynkins diagrams. The last  class  is the most studied one due to
its relation with  the physics of D-branes placed near to
singularities classified by Cartan matrices  of Lie algebras. In
this way, the physical content can be encoded in quivers sharing the
same properties like  Dynkin diagrams. A close inspection reveals
that one may have three models based on Lie algebras classification
with symmetric Cartan matrices \cite{ABS1,ABS2}.  The classification
of Dynkin diagrams led to the following  three different quiver
gauge theories:
\begin{enumerate}
\item Ordinary ADE quiver gauge theories
\item  Affine ADE quiver gauge theories
 \item Indefinite quiver gauge theories.
 \end{enumerate}

In string theory and related theories, the  models of first class
are not conformal invariant \cite{KMV}. They  can be obtained from
D-branes wrapping a collection of intersecting  cycles according to
ordinary ADE Dynkin diagrams of finite Lie algebras. The second
class of models are conformal gauge theories involving  a remarkable
realization in terms of D-branes wrapping elliptic singularities
\cite{KMV,BFS}. The last one is  the poor one from  the D-brane
realization point of view. There are  few models based on such Lie
algebras including hyperbolic cases \cite{ABS1,ABS2}.  It is worth
nothing that the last two models can be considered as possible
extension of ordinary ADE quiver gauge
  theories.  The derivation of such   models is based on the  usual philosophy one uses in the
building of the  Dynkin diagrams from the finite ones by adding
extra  nodes. In  the  quiver method, this can be understood  by
adding more gauge factors  and matter fields.  More precisely, The
extra nodes allow one to implement new physical constraints on gauge
factors and matter field contributions. In particular, the affine
node has been explored to engineer geometrically conformal models
which have a remarkable realization as D-brane world volume gauge
theories obtained from elliptic singularities.

On the basis of  this classification of  quiver gauge theories, it
is natural to think about  quiver gauge theories   based on other
algebra structures. In this paper we will be interested in Lie
superalgebras. Such symmetries have been extensively studied in
various contexts  \cite{LSA1,LSA2,LSA3,LSA4,LSA5,LSA6}. They can be
thought of as  a special extension of  bosonic  Lie symmetries.
Using quiver method, such symmetries can be explored to build a  new
class of quiver gauge theories. This may give an explicit evidence
for the role played by Lie superalgebras   string theory and related
models. It may also open  an issue to look for the corresponding
singularities in  the  compactification mechanism including
superCalabi-Yau manifolds.

Roughly speaking,   there are  many examples of such  Lie algebras.
In this paper we will be interested in the case  of $A(1,0)$ which
can be thought of as  a particular extension  of $A_1$ bosonic  Lie
algebras.  This example has been shown to deal with several aspects
of (supersymmetric) integrable conformal models \cite{SS1, SS2}. In
particular, it is  relevant in the classification program of
extended ${\mathcal{N}}= 2$ superconformal algebras and string
theories obtained by gauging ${\mathcal{N}}= 2$ Wess-Zumino-Witten
models. The choice of $A(1,0)$ is motivated from the fact that $A_1$
is considered as the building block of the ADE classification of
simply laced  Lie algebras. Indeed, since the algebra $A(1,0)$ is
the simplest extension of Lie superalgebras associated with  the ADE
classification, one can  shortly review  its structure. This
structure is quite the same as the case of $A_2$ but with some
novelties. Let us give briefly how this symmetry can be built from
the underlying bosonic one.   More details  can be fund in
\cite{Sorba}. Indeed, $A(1,0)$ is an eight (4+4) dimensional algebra
with rank 2. From the fundamental 3-dimensional representation of
$A(1,0)$, one can write down the relations obeyed by its four
bosonic generators and its four fermionic generators. Its root
system has been studied extensively involving  fermmionic and
bosonic roots. In fact,  it has been shown  that this algebra has
two different root systems.  The first root system involves  two
fermionic nonzero roots $\alpha_1, \alpha_2 $ having  length square
zero, and a normalized bosonic nonzero root with length square 2
given by  $(\alpha_1+\alpha_2) $. The corresponding Cartan matrix
reads as
\begin{equation}
K_{ij}=\left(
  \begin{array}{cc}
    0 & 1 \\
    1 & 0 \\
  \end{array}
\right) \label{matrix1}
\end{equation}
and it  is  associated with the following Dynkin diagram
\begin{figure}[tbph]
\centering
\begin{center}
\includegraphics[width=4cm]{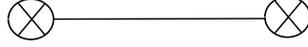} \caption{Dynkin diagram of A(1,0) associated with two fermionic simple
roots.
 }
\end{center}
\end{figure}

The second  possibility contains one  fermionic  simple nonzero root
$\alpha_1$  having length square zero and a simple   bosonic root
$\alpha_2 $ with length square 2.  The normalized fermionic one
nonzero root with length square 0 is  $(\alpha_2-\alpha_1) $. The
total root system is given by $\{\pm\alpha_1,
\pm\alpha_2,\pm(\alpha_2-\alpha_1)\}$. The corresponding Cartan
matrix takes the form
\begin{equation}
\label{matrix2}
 K_{ij}=\left(
  \begin{array}{cc}
    0 & 1 \\
    -1 & 2 \\
  \end{array}
\right)
\end{equation}
This matrix reproduces the following Dynkin Diagram
\begin{figure}[tbph]
\centering
\begin{center}
\includegraphics[width=4cm]{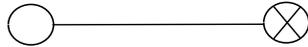} \caption{Dynkin diagram of A(1,0) associated with
  one  fermionic simple root and one bosonic simple root. }
\end{center}
\end{figure}

At this level some comments should  be done:  \begin{itemize}
 \item First,  we observe that  $A(1,0)$ is realized by two different Cartan
matrices leading to two different Dynkin diagrams. One of them is
purely fermionic.
 \item Second, these matrices are different from the one associated
 with the bosonic $A_2$  Lie algebras  given by
\begin{equation}
\label{matrix3}
 \left(
  \begin{array}{cc}
    2 & -1 \\
    -1 & 2 \\
  \end{array}
\right)
\end{equation}
\end{itemize}

Based on the correspondence between quiver gauge theories and Dynkin
diagrams, we would like  to construct quivers based on  Lie
superalgebras. This may offer a new approach  to deal with such
algebras from string theory point of view using geometric
engineering method. As usually, for each Dynkin diagram of A(1,0)
one associates a quiver model  which  can obtained from D-branes
wrapping appropriate cycles either in Calabi-Yau manifolds or
orbifolds spaces. The corresponding quiver gauge models constitute a
special class   based on  bosonic and fermionic nodes.

 To give a quiver gauge
theory associated with the above  Cartan matrices, it is interesting
first to find a geometric meaning of the
 the corresponding  roots in homology of  Calabi-Yau  Compactifications.
 Inspired from    bosonic quiver gauge theories, the above
 roots should  have a possible connection with middle cohomology groups  in  internal spaces.
 Based on geometric engineering method in
 II superstrings, the roots corresponding to  the above  Cartan
 matrices  may produce  a particular set of  intersecting cycles    embedded in two dimensional Calabi-Yau  manifolds
 including the
  K3 surfaces and ALE spaces.

  A close inspection in intersecting theory reveals that a possible
candidate can be given in terms of two intersecting homological two
cycles.   This  observation may  help us  to get  information  on
the complex two dimensional  homology of the possible  candidate for
the geometry associated with the Cartan matrices of A(1,0) Lie
superalgebra. A priori, they are many way one may use   to give a
consistent interpretation. However, the way we follow here   will be
based on the nice interplay  between toric geometry of the ADE
surfaces and Lie algebras \cite{KMV}. Indeed, the connection we are
after requires the introduction of complex cycles with genus $g$. In
fact, we will associate a simple root with a Riemann surface of
genus g that substitutes the $\mathbb{CP}^1$ sphere appearing in the
bosonic ADE classification.  To get our formulation,  we
use   the following points:\\
\begin{enumerate}
\item Supercommutator of $Z_2$ graded  satisfying the relation
\begin{equation*}
[x,y]=-(-1)^{|x||y|}[y,x]
 \end{equation*} where  de $|x|$ denotes
the degree of x either 0 or 1.
 \item The study of complex curves in two dimensional complex
surfaces.
\end{enumerate}
 Based on these points,  we  can associate
to a  root  system a set of complex curves according to the
following rules:
\begin{enumerate}
\item  one associates
to each bosonic  root  a complex curve of zero genus
\item one associates to each fermionic root  a complex curve with
genus one.
\end{enumerate}
These values of genus $(g=0,1)$ are inspired from the $Z_2$ graded
structure. With these ingredients, the  Cartan  matrices of A(1,0)
can be related to  a complex surface with rank 2 cohomology space
$H^2(X,Z)$. The later is generated by two cycle $C_i$ genus $g_i$.
More specifically, we propose the intersection form
\begin{equation}
\label{intersection}
C_iC_j=-2\delta_{ij}(g_i-1)+(-1)^{(g_i+g_j)}(1-\delta_{ij}) C_jC_i
\end{equation}
where $g_i=0,1$. This formulation has the following nice features.
First,  these intersection numbers reproduce the elements of the
above  Cartan matrices. Second, for $g_i=0$, we recover the bosonic
case with rank 2 given in (\ref{matrix3}). This formulation may be
considered as a complete picture for  the above three Cartan
matrices.

It is worth noting that this formulation is adaptable to all simply
laced superalgebras including $A(m,n)$ cases. We anticipate that the
evaluation of non simply laced in general will not be easy. This
will be addressed elsewhere where we also intend to discuss the
implementation of complex curves with higher genus. It would
therefore be of interest to try to extract information on graded
structure based on higher order discrete groups. We believe that
this  observation  deserves to be studied further.

As already mentioned, all what we know about   quiver gauge theories
can be extended to models based on Lie superalgebras. Mimicking
  the analysis one has done for the above  classification models, we
  can get a D-brane realization dealing with  quivers based on Lie
  superalgebras. The general study is beyond the scope of the present work, though we will consider two
explicit examples relying on   the above Cartan matrices. The gauge
symmetry  of such quivers involves naturally two  factors
 U$(N_1)\times$U$(N_2)$  with running gauge coupling constants $\lambda_1$,
$\lambda_2$ respectively. Each  factor  is engineered on   a node of
the Dynkin diagrams of  A(1,0) Lie superalgebra. The corresponding
models are quite similar to the ones built  on the ordinary $A_2$.
More precisely,  they   can be viewed as  a deformation  of $A_2$
quiver gauge theory  studied in \cite{KMV}. The only difference will
appear in the computation of the beta function. In $N = 2$
supersymmetric models in four dimensions \cite{KS,LNV}, the
holomorphic beta functions  $\beta_i$ depend linearly on the ranks
of the gauge factors where it has been shown to be proportional to
Cartan matrices. Taking into account this observation, the
calculation  can be extended to Lie superalgebras, namely
\begin{equation}
\label{beta} \beta_i\sim K_{ij}(g_i,g_j)N_j
\end{equation}
where now $K_{ij}(g_i,g_j)$ are Cartan matrices  of  the above Lie
superalgebras which can be  reproduced from (\ref{intersection}). In
$N=2$ quivers, the vanishing condition of such functions  is
intimately related to the conformal invariance \cite{KS,LNV}.  This
condition has a  nice interpretation in toric geometry realization
of Calabi-Yau manifolds \cite{F,LV}. In such a framework,
equation(\ref{beta}) can be reinterpreted as a relation between
toric vertices  defining the toric manifolds on which the quiver is
built. It has been observed that  the  conformal condition can be
reached by considering the following general relation
\begin{equation}
\label{matter}
 K_{ij}(g_i,g_j){ N}_j -M_i=0
\end{equation}
where $M_i$  are  fundamental matter fields required by the
Calabi-Yau constraint. It turns out that there are many ways to
solve such equations.  As the bosonic case,  a possible way  forces
us  to implement auxiliary nodes producing quiver  with more than
two gauge factors.  This procedure allows one to modify the above
Cartan matrices leading to a Calabi-Yau geometry. However,  to get
the desired quiver the extra nodes  behave like non dynamic gauge
factor and they should be associated with matter fields. In fact,
these models can be obtained by assuming that auxiliary nodes
represent non compact cycles in toric geometry realization of
Calabi-Yau manifolds\cite{KMV}.

Let us give concrete  examples. First we consider the case where
$g_i=1$ associated with (\ref{matrix1}). Indeed, mimicking the
analysis made for bosonic ADE  Lie algebras, the Calabi-Yau
condition requires the following  modified Cartan matrix
\begin{equation}
\widetilde{{K_{i\ell}}}(g_i=1)=\left(
  \begin{array}{cccc}
   -1& 0& 1 & 0 \\
    0 & 1&0&-1 \\
  \end{array}
\right) \label{matrix4}
\end{equation}
This matrix can have a  nice  physical interpretation   in  terms of
$N = 2$ linear sigma models  describing  the conformal field theory
on   Calabi-Yau string  background,  required by
$\sum_i\widetilde{{K_{i\ell}}}=0$. This target  string  background
can be analyzed by solving the D-term potential dealing with a gauge
group U(1)$^2$ with 4 matter fields $\psi_i$
 \bea
 \label{sigma}
 -|\psi_1|^2+|\psi_3|^2= r_1\nonumber \\
|\psi_2|^2-|\psi_4|^2= r_2, \eea where the $r_{1,2}$  are FI
coupling parameters.
 In this sigma model realization, the
U(1)$^2$ gauge symmetry can be identified with the Cartan subalgbras
of A(1,0) Lie superalgebra. Equation (\ref{sigma}) have a nice toric
geometry relation given by \be
 \widetilde{{K_{i\ell}}} v_\ell=0
 \label{toric}
 \ee
 where  $v_\ell$ toric  vertices are given by
 $v_\ell=(N_\ell,\star,\star)$
  required  by the Calabi-Yau conditions. In way, $N_\ell$ can be identified with   gauge group
  ranks.  However, the other components  are spectator  integers playing  no role in quiver
  models.

It turns out that equation (\ref{beta}) may be thought of as a
particular situation of the  equation (\ref{toric}). Inspired from
the trivalent vertex  method used in conformal field theory,  the
above toric equations   can be interpreted as quiver equations
associated with fundamental matter. Assuming that the  gauge
couplings associated with the external nodes go to zero, the
corresponding quiver is described by U$(N)\times$U$(N)$ gauge
symmetry and U$(N)\times$U$(N)$ flavor group. This quiver data may
be  obtained from D-branes wrapped on cycles with large volumes.
This model is encoded in figure 3 where the external nodes are
associated with U$(N)\times$U$(N)$ flavor group.
\begin{figure}[tbph] \centering
\begin{center}
\includegraphics[width=8cm]{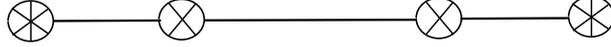} \caption{The middle nodes are associated with U$(N)\times$U$(N)$ gauge symmetry while the external
nodes are associated with fundamental matter }
\end{center}
\end{figure}

 To get the quiver associated the second Dynkin
diagram, one may  follow the  approach used for the first example.
Taking the corresponding modified Cartan matrix
\begin{equation}
\widetilde{{K_{i\ell}}}(g_1=1,g_2=0)=\left(
  \begin{array}{cccc}
   -1& 0& 1 & 0 \\
    0 & -1&2&-1 \\
  \end{array}
\right) \label{matrix5}
\end{equation}
required by conformal invariance in toric geometry language,  we
derive the quiver model associated with the Cartan matrix  given in
(\ref{matrix2}). This can  be obtained  by  considering the
vanishing limits of the gauge coupling constants associated with two
external nodes corresponding to the extra lines and columns of the
matrix (\ref{matrix5}). Equivalently, we take the extra  cycles to
be very large. The dynamics associated with such cycles become week
leading to a spectator symmetry group. This assumption produces a
quiver theory with  U$(N)\times$U$(N)$ gauge symmetry and flavor
group  of type U$(N)\times$U$(N)$ engineered on the two external
nodes.

\begin{figure}[tbph]
\centering
\begin{center}
\includegraphics[width=8cm]{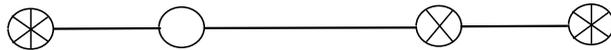} \caption{ Quiver model based on the second Cartan matrix of A(1,0). }
\end{center}
\end{figure}

In the end of this analysis, it is interesting to note  the
following crucial points:
\begin{enumerate}
\item A(1,0) Lie superalgebra encodes only a quiver gauge model even it
involves two Dynkin diagrams. We do not understand this coincidence
on the physical content of $A(1,0)$ Dynkin diagrams. We believe that
it should exist  a deeper reason  behind such a feature.  This
observation needs to be developed in future.

\item This analysis may be extended to the case of ADE superlagebras.
In this way,  one has nodes  with more than two links. This involves
polyvalent vertices in  toric geometry representation of Calabi-Yau
manifolds. More details on such  geometry can be found in
\cite{KMV}. In all geometries, the conformal condition may be
considered  as a particular expression  of toric geometry equations.

\item  We expect  that the  approach is adaptable to a broad variety of
geometries represented by   supermanifolds constructed as fermionic
extensions of  toric  Calabi-Yau varieties.  The geometries  can be
realized as target spaces of supersymmetric  linear sigma models
with  chiral and fermionic superfields with charges satisfying  the
super Calabi-Yau condition. A special class of such geometries with
ADE singularities have been discussed in \cite{BDRSS}. It should be
interesting to make contact with theses activities.
\end{enumerate}

As we have seen our work comes up with many questions. We try to
address elsewhere such issues.

\end{document}